\documentclass[12pt]{article}

\usepackage{amssymb}
\usepackage[dvips]{graphicx}
\usepackage[dvips]{psfrag}
\usepackage{cite}

\newcommand {\beq}{\begin{equation}}
\newcommand {\eeq}{\end{equation}}
\newcommand {\beqa}{\begin{eqnarray}}
\newcommand {\eeqa}{\end{eqnarray}}
\newcommand {\n}{\nonumber \\}

\begin{document}
\setlength{\oddsidemargin}{0cm}
\setlength{\baselineskip}{7mm}

\begin{titlepage}
\begin{normalsize}
\begin{flushright}
\begin{tabular}{l}
OU-HET 576\\
March 2007
\end{tabular}
\end{flushright}
  \end{normalsize}

~~\\

\vspace*{0cm}
    \begin{Large}
       \begin{center}
         {On-shell Action for Bubbling Geometries}
       \end{center}
    \end{Large}
\vspace{1cm}

\begin{center}
           Matsuo S{\sc ato}\footnote
            {
e-mail address : 
msato@het.phys.sci.osaka-u.ac.jp}\\
           
      \vspace{1cm}
       
        {\it Department of Physics, Graduate School of  
                     Science}\\
               {\it Osaka University, Toyonaka, Osaka 560-0043, Japan}\\
    \end{center}

\hspace{5cm}

\begin{abstract}
\noindent

We study the half-BPS sector of the AdS/CFT correspondence. In the full sector of the correspondence, on-shell actions in type IIB supergravity have been played crucial roles, for example in the GKP-Witten relation. We derive, therefore, an on-shell action that reproduces a class of half-BPS solutions found by Lin, Lunin and Maldacena. These solutions describe the bubbling geometries in the supergravity. We show that this on-shell action for the LLM solutions is dual to the classical limit of an observable of the N free fermions in a harmonic oscillator potential in 1+1 dimensions. The observable consists of three parts: an expectation value of the total energy, matrix elements of the momentum squared, and matrix elements of the position squared. Our results suggest that there exists a holographic correspondence on a two-dimensional surface in the bubbling geometries.


\end{abstract}

\vfill
\end{titlepage}
\vfil\eject

\setcounter{footnote}{0}

\section{Introduction}
\setcounter{equation}{0}

The AdS/CFT correspondence has attracted much interest in the last decade. In a convincing part of the AdS/CFT correspondence, there exists a holographic correspondence between a five-dimensional supergravity around the $AdS_5$ background that is reduced from ten-dimensional type IIB supergravity on $S^5$ and four-dimensional large N $\mathcal{N}=4$ super Yang-Mills (SYM) living on the AdS boundary \cite{Maldacena}. In this correspondence, the on-shell actions in the supergravity have been played crucial roles \cite{GKP, Witten, ST}. Especially, the on-shell action of the AdS supergravity coincides with the generating functional of the correlation functions in the SYM. In this relation, which is called the Gubser Klebanov Polyakov - Witten (GKP-Witten) relation \cite{GKP, Witten}, the boundary values of the fields on the AdS boundary in the supergravity are identified with source functions that couple to the operators in the SYM.  

Recently, significant advances have been made in the half-BPS sector of the AdS/CFT correspondence \cite{CorleyJevickiRamgoolam, Berenstein, LinLuninMaldacena}. In this sector, the full correspondence is reduced to a correspondence between a class of half-BPS solutions found by Lin, Lunin and Maldacena in ten-dimensional type IIB supergravity, and N free fermions in a harmonic oscillator potential in 1+1 dimensions. We call the solutions that belong to this class the Lin-Lunin-Maldacena (LLM) solutions. The LLM solutions describe bubbling geometries, which include $AdS_5 \times S^5$ geometry and giant gravitons. These solutions are half-BPS solutions with $SO(4) \times SO(4) \times R$ isometry. Under these conditions, the equations of motion of the type IIB supergravity are consistently reduced to a single three-dimensional equation of motion. The consistent reduction means that all solutions of the three-dimensional equation can be lifted to solutions of the type IIB supergravity. Each LLM solution is determined by solving the three-dimensional equation of motion and fixing boundary values on a two-dimensional boundary. If we assume that the LLM solutions are non-singular, the boundary values are restricted to take only 0 or 1, and thus they form droplets on the two-dimensional boundary. On the other hand, in this sector, the SYM is reduced to a system that consists of N free fermions in a harmonic oscillator potential in 1+1 dimensions. Each state of this system is represented by certain fermion droplets on the two-dimensional phase space. In this correspondence, the boundary of the three-dimensional space is identified with the phase space of the fermions. Moreover, the boundary values of the scalar are identified with the droplets of the fermions on the phase space \cite{LinLuninMaldacena, Wigner, DasDharMandalWadia, Mandal, TakayamaTsuchiya}. Thus, it is natural to expect that there exists a holographic correspondence between the three-dimensional scalar and the fermions on the two-dimensional phase space as in the case of the full sector.
 
In this paper, we study an on-shell action for the three-dimensional scalar and its dual in order to clarify the holographic correspondence between the three-dimensional scalar and the fermions on the two-dimensional phase space. We study other holographic correspondence on the two-dimensional surface than the holographic correspondence on the AdS boundary in the full sector. Actually, the two-dimensional boundary is not a part of the AdS boundary \footnote{Authors in \cite{JevickiYoneya, Tai} discuss how the GKP-Witten relation on the AdS boundary is reduced in the half-BPS sector.}. In section 2, we shortly review on the half-BPS sector of the AdS/CFT correspondence. In section 3, we derive an on-shell action for the LLM solutions analogously to the GKP-Witten case. We show that our on-shell action one-to-one corresponds to the class of the LLM solutions. The precise description of the one-to-one correspondence is given in this section. In section 4, by using the relation between the boundary values of the scalar and the fermion droplets, we show that the on-shell action for the LLM solutions is dual to the classical limit of an observable of the N free fermions in the harmonic oscillator potential in 1+1 dimensions. The observable consists of three parts: an expectation value of the total energy, matrix elements of the momentum squared, and matrix elements of the position squared. In section 5, we summarize and discuss the results. 






\vspace{1cm}

\section{Review on Half-BPS AdS/CFT Correspondence}
\setcounter{equation}{0}

In this section, we briefly review on the half-BPS sector of the AdS/CFT correspondence \cite{Berenstein, LinLuninMaldacena}. In this sector, the AdS/CFT correspondence is reduced to a correspondence between a class of the LLM solutions on the gravity side and N free fermions in a harmonic oscillator potential in 1+1 dimensions on the gauge theory side. 

On the gravity side, the LLM solutions are half-BPS solutions with $SO(4) \times SO(4) \times R$ isometry in ten-dimensional type IIB supergravity,
\begin{eqnarray}
&& ds^2 = -h^{-2}\left[dt + V\right]^2 
+ h^2 \left[dy^2 + dx^idx^i\right] + y e^{G}d\Omega^2_3 + y e^{-G}d\widetilde{\Omega}^2_3\,, \label{LLMmetric} \qquad \n
&&F_{(5)}=F_{\mu\nu}dx^{\mu}\wedge dx^{\nu} \wedge d\Omega_3
+\tilde{F}_{\mu\nu}dx^{\mu} \wedge dx^{\nu} \wedge d \tilde{\Omega}_3, \n
&& \qquad 
 h^{-2} = 2y\cosh G\,, \quad z= \frac{1}{2}\tanh G\,,  \quad V=V_{i}dx^i\,,
\nonumber \\ 
&& \qquad y \partial_y V_i = \epsilon_{ij} \partial_j z \,, \quad 
y(\partial_iV_j-\partial_jV_i) = \epsilon_{ij}\partial_yz\,, \label{z-v} \n
&& \qquad F = dB_t\wedge (dt+V)+B_tdV + d\widehat{B}\,, \quad 
 \widetilde{F} = d\widetilde{B}_t\wedge(dt+V) + \widetilde{B}_t dV 
+d\widehat{\widetilde{B}}\,, \nonumber \\
&& \qquad 
B_t = - \frac{1}{4}y^2 e^{2G}\,, \quad \widetilde{B}_t 
= - \frac{1}{4}y^2 e^{-2G}\,,\nonumber \\ 
&& \qquad  
d\widehat{B} = -\frac{1}{4}y^3\ast_3d 
\left(\frac{z+\frac{1}{2}}{y^2}\right)\,, 
\quad d\widehat{\widetilde{B}} = - \frac{1}{4}y^3\ast_3 d\left(
\frac{z-\frac{1}{2}}{y^2}\right)\,, \label{ansatz}
\end{eqnarray}
where $i, j$ run 1, 2, $\mu, \nu$ run $0, \cdots, 3$, $x^0=t$ and $x^3=y$. $*_3$ represents a Hodge dual in the three dimensions, parametrized by $x^1, x^2, y$. In this solution, the dilaton and axion are constant and the three-form field strengths are zero. This solution is determined by a single function $z$, which is a solution of a three-dimensional second-order partial differential equation,
\begin{eqnarray}
\partial_i\partial_i z + y\partial_y\left(\frac{\partial_y z}{y}\right) = 0.
\label{eom} 
\end{eqnarray} 
(\ref{eom}) is a consistent reduction from ten-dimensional type IIB supergravity under the ansatz (\ref{ansatz}). The consistent reduction means that all solutions of (\ref{eom}) can be lifted to solutions of the equations of motion of ten-dimensional type IIB supergravity by using (\ref{ansatz}). That is, this three-dimensional equation of motion gives dynamics on the supergravity side. Actually, the LLM solutions are determined by a function, which are solutions of (\ref{eom}) with Dirichlet boundary conditions $z(x_1, x_2, 0)$ on a two-dimensional surface specified by $y=0$,
\beq
z_{LLM}(x_1, x_2, y)=\frac{1}{\pi}\int_{\mathcal{D}}dx_1'dx_2'\frac{y^2}{[(x_1-x_1')^2+(x_2-x_2')^2+y^2]^2}z(x_1', x_2', 0). \label{LLM}
\eeq
If we assume that the LLM solutions are non-singular, $z(x_1, x_2, 0)$ are restricted to take only 1/2 or -1/2. As a result, $w(x_1, x_2)=z(x_1, x_2, 0)+\frac{1}{2}$ take only 0 or 1, and thus they form droplets.

On the gauge theory side, large N $\mathcal{N}=4$ super Yang-Mills (SYM) is reduced to a system consisting of N free fermions in a harmonic oscillator potential in 1+1 dimensions in the half-BPS sector. On a phase space, a state of fermions can be expressed by fermion droplets. 

\begin{figure}[htbp]
\begin{center}
\psfrag{x1}{$x_1 \equiv p$}
\psfrag{x2}{$x_2 \equiv q$}
\psfrag{0}{$y=0$}
\psfrag{y}{$y$}
\includegraphics[height=4cm, keepaspectratio, clip]{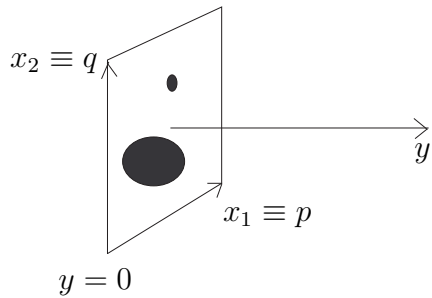}\end{center}
\caption{holography}
\label{holography}
\end{figure}

In this correspondence, the boundary plane $(x_1, x_2)$ in the three dimensions on the gravity side is regarded as the two-dimensional phase space $(p, q)$ of the fermions. Especially, the boundary values $w(x_1, x_2)$ of the scalar are identified with the fermion droplets $w(p, q)$ in (\ref{wigner}) as shown in Fig. \ref{holography}.  

 


\vspace*{1cm}

\section{On-shell Action for LLM Solutions}
\setcounter{equation}{0}
In the full AdS/CFT correspondence, the on-shell action around the $AdS_5 \times S^5$ background in the type IIB supergravity corresponds to the generating functional of correlation functions in large N $\mathcal{N}=4$ SYM (GKP-Witten relation \cite{GKP, Witten}). In this section, we derive an on-shell action for the LLM solutions analogously to the GKP-Witten case. 

Let us recall the GKP-Witten relation in the full sector of the AdS/CFT correspondence. On the gravity side, the AdS boundary is given by $r \to \infty$, where $r$ is the radial coordinate of the Poincare coordinates. We introduce as a cut-off another boundary that is given by $r=r_0$ ($r_0$ is finite). We define boundary values of fields on this cut-off boundary and obtain an on-shell action. Some terms in the on-shell action keep finite and the others diverge or vanish when the cut-off boundary approaches the AdS boundary. The finite terms give correlation functions in the large N $\mathcal{N}=4$ SYM (For a review, see \cite{AGMOO-DF}.).

We study an on-shell action in the half-BPS sector analogously to the full case. We begin with an action 
\beqa
  I=\int d^2x dy \frac{1}{2y} ((\partial_i z)^2 + (\partial_y z)^2), 
\label{action}
\eeqa
from which the equation of motion  (\ref{eom}) is derived. Actually, there are other actions from which the equation of motion  (\ref{eom}) is derived. In most cases, if one substitutes ansatz for a reduction to an action, one obtains a reduced action from which consistently reduced equations of motion are derived. However, in the type IIB supergravity case, we cannot derive consistently reduced equations of motion from the action which is obtained by substituting ansatz to an action, because of the self-duality condition. There is no unique way to obtain a reduced action. Therefore, we choose the most simple one (\ref{action}).

\begin{figure}[htbp]
\begin{center}
\psfrag{x1}{$x_1$}
\psfrag{x2}{$x_2$}
\psfrag{0}{$y=0$}
\psfrag{z}{$y=y_0$}
\psfrag{y}{$y$}
\includegraphics[height=4cm, keepaspectratio, clip]{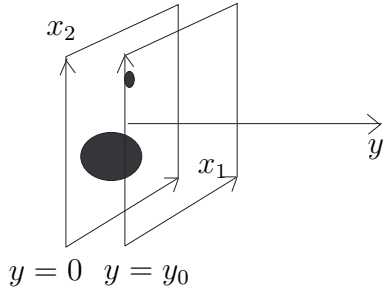}\end{center}
\caption{cut-off boundary}
\label{cut-off}
\end{figure}

The Hamiltonian for this action is given by
\beq
H=\int d^2x(\frac{1}{2}yp^2-\frac{1}{2y}(\partial_iz)^2).
\eeq
In a similar way to the full case, we introduce a cut-off boundary on $y=y_0$ as shown in Fig. \ref{cut-off}. By regarding $y$ as time and setting boundary values as $u(x_1, x_2):= z(x_1, x_2, y=y_0)$ on $y=y_0$, an on-shell action $S(u, y_0)$ for (\ref{action}) is defined. $S(u, y_0)$ satisfies the Hamilton-Jacobi (H-J) equation, 
\beq
\frac{\partial S}{\partial y_0} + H=0,
\eeq
where
momentum $p$ is replaced as 
\beq
p=\frac{\delta S}{\delta u}.
\eeq
Explicitly, the H-J equation is expressed as
\beqa
   \frac{\partial S}{\partial y_0} +\int d^2 x (\frac{1}{2} y_0 
(\frac{\delta S}{\delta u})^2) + \frac{1}{2y_0} \int d^2 x u 
(\partial_1^2+\partial_2^2) u =0.       \label{HJ}  
\eeqa

It is difficult to solve (\ref{HJ}). Instead, in order to obtain solutions of (\ref{HJ}), we solve (\ref{eom}) with Dirichlet boundary conditions and substitute the solutions to the action. We set the following boundary conditions under which the solutions reduce to (\ref{LLM}) when $y_0 \to 0$: 
\beqa
&&z(x_1, x_2, y_0)=u(x_1, x_2) \n
&&z(\pm \infty, x_2, y)=z(x_1, \pm \infty, y) = z(x_1, x_2,  \infty)=0.
\label{boundarycondition}
\eeqa

By defining $f(k_1, k_2, t)$ by
\beq
f(k_1, k_2, t)
=
\int d x_1 d x_2 e^{i(k_1 x_1 +k_2 x_2)}\frac{k}{t} z(x_1, x_2, \frac{t}{k}),
\eeq
(\ref{eom}) is reduced to the modified Bessel differential equation,
\beq
(\partial_t^2 +\frac{1}{t} \partial_t -(1+\frac{1}{t^2}))f=0.
\eeq     
(\ref{boundarycondition}) selects one of the two independent solutions, the modified Bessel function of the second kind $K_1(t)$ and determines the coefficient. As a result, we obtain 
\beqa
z(x_1, x_2, y; y_0)
&=&
\frac{y}{y_0}\int d^2x'  u(x') \frac{1}{(2\pi)^2} \int d^2k
e^{ik_1(x_1-x_1')+ik_2(x_2-x_2')}
\frac{K_1(ky)}{K_1(ky_0)}\n 
&=&
\frac{y}{y_0}\int d^2x' u(x') 
\int_0^{\infty}dk k I_0(ik|x-x'|)
\frac{K_1(ky)}{K_1(ky_0)} \label{solution},
\eeqa
where $I_0$ is the modified Bessel function of the first kind. This solution satisfies (\ref{eom}) and (\ref{boundarycondition}).

These solutions one-to-one correspond to (\ref{LLM}). Actually,
\beqa
\lim_{y_0 \to 0}z(x_1, x_2, y; y_0)
&=&
\frac{y}{2\pi} \int d^2x' u(x') \int_0^\infty dk k^2 I_0(ik|x-x'|)K_1(ky) \n
&=&
\frac{y}{2\pi} \int d^2x' u(x')\frac{\pi}{2y^3}\frac{4\Gamma(2)\Gamma(1)}{\pi\Gamma(1)}F(2,1,1;(-i\frac{x}{y})^2) \n
&=&
\frac{y}{2\pi} \int d^2x' u(x') \frac{2}{y^3}\frac{1}{(1+(\frac{x}{y})^2)^2} \n
&=&
\frac{1}{\pi} \int d^2x' \frac{y^2}{((x_1-x_1')^2+(x_2-x_2')^2 +y^2)^2} u(x')\n
&=&
z_{LLM}(x_1, x_2, y),
\eeqa
where $F(a,b,c;x)$ is the Gauss's hypergeometric function.

If we substitute the solutions (\ref{solution}) to the action (\ref{action}), only surface terms on $y=y_0$ survive. We obtain 
\beqa  
S&=& \frac{1}{2y_0^2} \int d^2x u^2(x) \n
&+& \frac{1}{4 \pi y_0}
\int d^2x d^2x' u(x) u(x') \int_0^{\infty} dk k^2 
I_0(ik|x-x'|) \frac{K_1'(y_0 k)}{K_1(y_0 k)},              
\eeqa
where $K_1'(y)= \frac{\partial K_1(y)}{\partial y}.$ 
Actually, we can easily verify that this satisfies the H-J equation (\ref{HJ}).
This on-shell action is uniquely determined by the LLM solutions because the on-shell action is obtained by substituting (\ref{solution}), which one-to-one correspond to (\ref{LLM}). On the other hand, the on-shell action reproduces (\ref{LLM}) and thus the LLM solutions, because the on-shell action satisfies the H-J equation. That is, the on-shell action determines a first order differential equation that is obtained by integrating the equation of motion (\ref{eom}), and (\ref{solution}) satisfies this first order differential equation. Therefore, the on-shell action one-to-one corresponds to the class of the LLM solutions.


In a similar way to the full sector, it is natural to expect that finite parts of the on-shell action give physical quantities on the gauge theory side also in the half-BPS sector, when the cut-off boundary approaches the boundary. Therefore, we need to expand the on-shell action in $y_0$ around $y_0=0$. Non-vanishing terms in $y_0 \to 0$ are given by 
\beqa
\lim_{y_0 \to 0}S
&=& -\ln(y_0) \frac{1}{2}\int d^2 x u (\partial_1^2+\partial_2^2) u \n
&&+\frac{\ln 2-\gamma}{2}\int d^2 x u (\partial_1^2+\partial_2^2) u
-\frac{1}{4}\int d^2 x u (\partial_1^2+\partial_2^2) \ln(-\partial_1^2-\partial_2^2) u, \label{y=0}
\eeqa
where $\gamma$ is Euler constant. The local part of the constant terms of the on-shell action is 
\beq
\frac{\ln 2-\gamma}{2} \int d^2 x u (\partial_1^2+\partial_2^2) u. \label{on-shell}
\eeq
We study what is dual to (\ref{on-shell}) in the next section. 

\vspace{1cm}

\section{Holographic Dual to On-shell Action}
\setcounter{equation}{0}
In this section, we study what is holographic dual to the on-shell action (\ref{on-shell}). On the gauge theory side, in order to express the fermion droplets, we use the Wigner distribution function \cite{Wigner} given by, 
\beqa
\hat{w}(p, q)=\frac{1}{\hbar}\int dx e^{ipx/\hbar} \psi^{\dagger}(q+x/2) \psi(q-x/2), \label{wigner}
\eeqa
where $\psi$ is a wavefunction operator of second quantized fermions. $\hat{w}(p,q)$ satisfies the following properties,
\beqa
&&\int dp dq \, \hat{w}(p,q)=N  \n 
&&\frac{\partial}{\partial t}\hat{w}
= (q \frac{\partial}{\partial p}
-p\frac{\partial}{\partial q})\hat{w}  \n 
&&\hat{w}(p,q)\ast \hat{w}(p,q)= \hat{w}(p,q),
\eeqa  
where $\ast$ is a Moyal product.
In the classical limit, $\hat{w}(p,q)$ satisfies
\beqa
&&\int dp dq \, \hat{w}(p,q)=N  \n 
&&\frac{\partial}{\partial t}\hat{w} 
= (q \frac{\partial}{\partial p}
-p\frac{\partial}{\partial q})\hat{w}  \n 
&&\hat{w}(p,q)^2 = \hat{w}(p,q) \leftrightarrow  \hat{w}(p,q)=0,1. \label{conditionfordroplets}
\eeqa  
For energy eigen states, fermion droplets $w(p,q)$ can be defined \cite{Wigner, DasDharMandalWadia, TakayamaTsuchiya} by
\beqa
w(p, q)=\lim_{\hbar \to 0} \langle \hat{w}(p,q) \rangle, \label{wigner}
\eeqa
where $\rangle$ is an energy eigen state of the fermions. $w(p, q)$ satisfies the same properties as in (\ref{conditionfordroplets}), which are conditions for the droplets \footnote{The expectation values of the Wigner distribution function for the other states do not satisfy the condition for the droplets. This is a limitation of our methods. We need another formulation for the droplets in order to study the other states.}. This formula describes a relation between a shape of the droplets and the state of the fermions. 

The fermionic field can be expanded as
\beq
\psi(x,t)=\sum_{n=0}^{\infty} C_n e^{-i E_n t/\hbar }\psi_n(x),
\eeq
where
\beqa
&&H \psi_n(x)=E_n\psi_n(x) \n
&&\{C_m, C_n^{\dagger} \}=\delta_{m,n} \n
&&\sum_{n=0}^{\infty}C_n^{\dagger} C_n=N.
\eeqa
We do not assume any form of the Hamiltonian. The energy eigen states are given by 
\beqa
\mid \rangle
&=&
\mid n_1, n_2, \cdots, n_N \rangle \n
&=&
C^{\dagger}_{n_1} C^{\dagger}_{n_2} \cdots C^{\dagger}_{n_N} \mid 0 \rangle, 
\eeqa
where
\beq
C_n \mid 0 \rangle=0.
\eeq
As a result, (\ref{wigner}) is reduced to 
\beqa
 w(p,q)= \lim_{\hbar \to 0}\sum_{i=1}^N \frac{1}{\hbar}\int dx e^{ipx/\hbar} \psi^{\dagger}_{n_i}(q+x/2)\psi_{n_i}(q-x/2).   
\eeqa

Finally, we identify the boundary values $u(x_1, x_2)+\frac{1}{2}$ with the expectation values of the Wigner distribution function $w(p, q)$. By substituting this form to (\ref{on-shell}), we obtain 
\beqa
 S_0&=& \lim_{\hbar \to 0} 
\Biggl[\frac{1}{2} \sum_{i=1}^N \frac{1}{\hbar^3} \int dx  \psi_{n_i}^{\dagger}(x)((i \hbar\partial_x)^2
+x^2) \psi_{n_i} (x) \n
&&- \frac{1}{2} \sum_{i,j=1}^N \frac{1}{\hbar^3} \Biggl( | \int dx  
\psi_{n_i}^{\dagger}(x) i \hbar \partial_x  \psi_{n_j} (x)|^2
+|\int dx
\psi_{n_i}^{\dagger}(x) x  \psi_{n_j} (x)|^2 \Biggr)\Biggr]. \label{dual}
\eeqa
The first term represents the total energy of N free fermions in the harmonic oscillator potential in 1+1 dimensions. This is a non-trivial result because we have not assumed any form of the Hamiltonian of the fermions. Final result shows that the on-shell action for the LLM solutions is dual to the classical limit of an observable of the N free fermions in the harmonic oscillator potential in 1+1 dimensions. The observable consists of three parts: an expectation value of the total energy, matrix elements of the momentum squared, and matrix elements of the position squared. 

\begin{figure}[htbp]
\begin{center}
\psfrag{a}{$AdS^5 \times S^5$}
\psfrag{b}{dual giant gravitons}
\psfrag{c}{giant gravitons }
\psfrag{d}{uniformly distributed}
\includegraphics[height=4cm, keepaspectratio, clip]{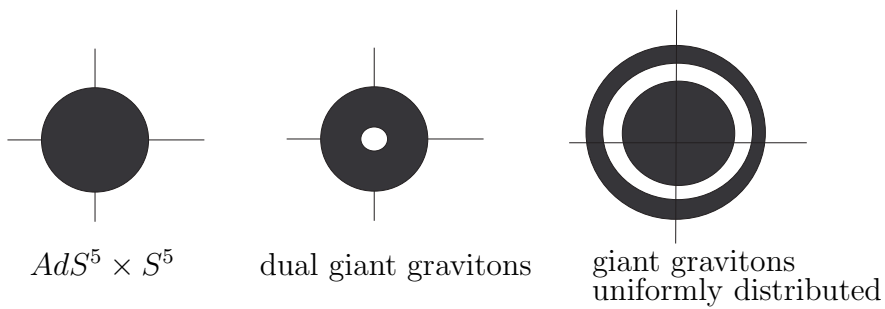}\end{center}
\caption{Boundary conditions corresponding to energy eigen states. Boundary values take 0 or 1 on white or black regions respectively.}
\label{droplets}
\end{figure}

When we discuss dual to the on-shell action, we consider the boundary conditions corresponding to the energy eigen states. Because we obtain the N free fermions in the harmonic oscillator potential in 1+1 dimensions, the energy eigen states represent $AdS^5 \times S^5$, dual giant gravitons in the $AdS_5 \times S^5$ background or giant gravitons distributed uniformly along an orbit on the $S^5$ in the $AdS_5 \times S^5$ background \cite{Berenstein, LinLuninMaldacena, TakayamaTsuchiya}, with respect to the boundary conditions as shown in Fig. \ref{droplets}.  

\vspace{1cm}

\section{Conclusion and Discussion}
\setcounter{equation}{0}

We have obtained an on-shell action for a scalar theory that is dimensionally reduced from ten-dimensional type IIB supergravity. This on-shell action one-to-one corresponds to a class of the LLM solutions, which represents the half-BPS sector of the AdS/CFT correspondence. We have shown that the on-shell action for the LLM solutions is dual to the classical limit of an observable of the N free fermions in the harmonic oscillator potential in 1+1 dimensions. The observable consists of three parts: an expectation value of the total energy, matrix elements of the momentum squared, and matrix elements of the position squared.  Our results imply that there exists a correspondence between the three-dimensional scalar and the fermions on a two-dimensional surface. This correspondence is different from the correspondence on the AdS boundary in the full sector. Therefore, our result should be a new aspect of the AdS/CFT correspondence.

There are some directions to study further. First, it is interesting to study what is dual to the non-local part of the finite terms in the on-shell action. In the full sector, such non-local parts give correlation functions on the gauge theory side. Second, we need another formalism for droplets on the phase space to study all fermionic states. The expectation values of the Wigner distribution functions only for the energy eigen states can form droplets on the phase space. Third, we need to deeply understand why the rather complicated quantity (\ref{dual}) on the gauge theory side is dual to the on-shell action for the LLM solutions on the gravity side.

We comment on an application of our results. The relation between the on-shell action and the fermionic observables should be extended to a half-BPS sector of the AdS/CFT correspondence where quantum corrections are included on the gravity side.  It is expected that we can rather easily treat quantum corrections on the gravity side in the half-BPS sector of the AdS/CFT correspondence. Actually, it is discussed that we only need to fluctuate the droplets in order to quantize the half-BPS sector of the type IIB supergravity, by using the Crnkovic-Witten-Zuckerman's method of mini-superspace quantization \cite{GrantMaozMarsanoPapadodimasRychkov}. Therefore, we expect that we can study quantum corrections to the LLM solution by studying only quantum corrections to the three-dimensional scalar. We need to find a wave function that reduces to the exponential of our on-shell action in the classical limit. Therefore, this study should be useful for an extension of the AdS/CFT correspondence where quantum corrections are included on the gravity side.

\vspace*{1cm}

\section*{Acknowledgements}
I would like to thank M. Fukuma, Y. Hyakutake, H. Kawai, I. Kishimoto, T. Takayanagi, K. Ohashi, K. Ohta, K. Okamura, and K. Yoshida for stimulating discussions. I am also grateful to A. Tsuchiya not only for stimulating discussions but also for useful comments on the manuscript. I also thank the Yukawa Institute for Theoretical Physics at Kyoto University. Discussions during the YITP workshop YITP-W-06-16 on "Fundamental Problems and Applications of Quantum Field Theory" were useful to complete this work. This work is supported in part by The 21st Century COE Program "Towards a New Basic Science; Depth and Synthesis.''

%

\end{document}